\documentclass[prl,preprint,superscriptaddress]{revtex4-1}
\usepackage{amsmath}
\usepackage{amsfonts,amssymb}
\usepackage{graphics,epsfig}
\usepackage{color}
\usepackage{colortbl}

\newcommand{\be}{\begin{equation}}
\newcommand{\ee}{\end{equation}}
\newcommand{\beq}{\begin{equation}}
\newcommand{\eeq}{\end{equation}}
\newcommand{\bea}{\begin{eqnarray}}
\newcommand{\eea}{\end{eqnarray}}
\newcommand{\ba}{\begin{eqnarray}}
\newcommand{\ea}{\end{eqnarray}}

\newcommand{\la}[1]{\label{#1}}

\def\one{\mbox{1 \kern-.59em {\rm l}}}

\begin{document}

\title{Universal Landau Pole}
\author{A.A.~Andrianov}
\email{andrianov@icc.ub.edu} \affiliation{Department of
Theoretical Physics, Saint-Petersburg State University, St.
Petersburg, Russia} \affiliation{Dept.
Estructura i Constituents de la Mat\`eria, and Institut de Ci\`encies del
Cosmos (ICCUB), Universitat de Barcelona, Barcelona, Catalonia, Spain}
\author{D. Espriu}
\email{espriu@icc.ub.edu}
\affiliation{Dept.
Estructura i Constituents de la Mat\`eria, and Institut de Ci\`encies del
Cosmos (ICCUB), Universitat de Barcelona, Barcelona, Catalonia, Spain}
\author{M.A. Kurkov}
\email{kurkov@na.infn.it}
\affiliation{Dipartimento di Fisica, Universit\`{a} di Napoli {\sl Federico II}, Napoli,
Italia}
\affiliation{INFN, Sezione di Napoli}
\author{F. Lizzi}
\email{fedele.lizzi@na.infn.it}
\affiliation{Dept.
Estructura i Constituents de la Mat\`eria, and Institut de Ci\`encies del
Cosmos (ICCUB), Universitat de Barcelona, Barcelona, Catalonia, Spain}
\affiliation{Dipartimento di Fisica, Universit\`{a} di Napoli {\sl Federico II}, Napoli,
Italia}
\affiliation{INFN, Sezione di Napoli}


\begin{abstract}
Our understanding of quantum gravity suggests that at the Planck scale the usual geometry loses its meaning.
If so, the quest for grand unification in a large non-abelian group naturally endowed 
with the property of asymptotic freedom may also lose its motivation.
Instead we propose a unification of all fundamental interactions at the Planck scale in the form of a 
\emph{Universal Landau Pole} (ULP), at which all gauge couplings diverge. The Higgs quartic 
coupling also diverges while the Yukawa couplings vanish. The unification is achieved with the addition 
of fermions with vector gauge couplings coming in multiplets and with hypercharges identical to those of 
the Standard Model. The presence of these particles also prevents the Higgs quartic coupling from becoming 
negative, thus avoiding the instability (or metastability) of the Standard Model vacuum.
\end{abstract}

\preprint{ICCUB-13-045}

\maketitle

Under the renormalization group flow the coupling constants of the three fundamental gauge interactions behave quite differently~\cite{particledata}.
While the couplings of the non-abelian interactions, weak and strong,
constantly diminish with as the energy increases, the coupling of the abelian interaction grows, 
and eventually diverges,
a phenomenon usually referred to as \emph{Landau pole}~\cite{landaupole}.

This results from a direct extrapolation of physics at present energies. The existence
of new particles, or in general new physics, alter this behavior. For some time it was thought  
that the three interactions
coincided at a single scale, and this was interpreted to
signal the presence of a non-abelian grand unified group. Present accurate data show
that this triple coincidence does not happen in the absence of new
physics. 
The weak and hypercharge couplings are equal at a scale
of the order of $10^{12}$~GeV, hypercharge and strong at around $10^{14}$~GeV and
the two non-abelian couplings meet around $10^{17}$~GeV.

Continuing the flow of the three couplings beyond the previous scales would give rise
to a "weak" force actually stronger than the "strong" one, and the abelian coupling overtaking both of them. 
If one continues the running of the hypercharge coupling it diverges at the finite, albeit
extremely large, scale of $10^{53}$~GeV. Even though the calculations are done perturbatively 
at a finite loop order and the value of the scale where the pole occurs is therefore not to be trusted 
numerically  (since the perturbative expansion will have broken before), the qualitative behavior 
will however remain: for a nonasymptotic free theory at some energy there will be a pole.

Also relevant is the behavior under the renormalization flow of the quartic Higgs coupling $\lambda$
and of the Yukawa couplings of the top $y_t$, which is the largest of the fermion couplings
and therefore dominates at high energy. They both decrease, but while $y_t$ remains positive, $\lambda$
becomes negative in the presence of a relatively light Higgs boson with mass around $\sim$125~GeV~\cite{HiggsMass}.
This signals an instability, or at least a metastable phase, of the theory.

The idea of the unification of the forces is very appealing. A grand unified group
guarantees the presence of asymptotic freedom and consequently the ability to describe particles
and fields at arbitrarily small distances. However, we know that in nature there is also gravity
and before the Planck scale, around $m_p=10^{19}$~GeV, the onset of quantum gravity
will certainly alter the picture in a substantial way. Models of emergent gravity
(see e.g. \cite{AlfaroEspriuPuigdomenech}) indicate that there may be a `smallest distance' below 
which the very notion of length may not exist. In any case, dramatic
quantum gravity effects ---perhaps a string theory--- are likely to manifest themselves at
around the Planck scale and it is not obvious at all why one should expect quantum field
theory to remain perturbatively valid at or beyond the Planck scale. Then, the philosophical 
necessity for asymptotic freedom at the most fundamental scale weakens considerably.

In this letter we want to put forward another type of unification. Namely, the proposal that all
coupling constants, as well as the Yukawa couplings and the quartic Higgs coupling have a
singularity at an energy of the order of the Planck mass. This common singularity, which we
term \emph{Universal Landau Pole} (ULP) may be interpreted as signaling the onset 
of a phase transition to radically new physics. 
The nature of the ``high energy'' trans-planckian phase is not known, there would probably be some 
sort of quantum space-time and hypotheses abound. The existence of a 
common singularity might hint that this
new phase could be weakly coupled, but in a completely different set of variables. We  
assume that singularity at the transition shows up as a pole in \emph{all} gauge 
couplings and the quartic coupling, and a zero for the Yukawa coupling.
In the following we will see that the model presented here also solves the potentially disastrous
instability of the Higgs potential~\cite{instability}. 

The recent Large Hadron Collider measurement~\cite{HiggsMass}
of the Higgs mass around $125 \pm 1$~GeV together with the absence of new physics indicate that the
quartic coupling of the Higgs self-interaction may become negative at an energy as low 
as $10^8$~GeV~\cite{EliasMiro:2011aa} suggesting an instability of the theory. 
In terms of the effective potential this is tantamount to a negative quadric term, and therefore 
the potential is not bounded from below.

In the scenario we propose, the presence of new particles solves the
stability problem, and at the same time generates the ULP. In order to achieve this
the new physics must hasten the running of the abelian coupling towards the pole, lowering it 
from $\sim 10^{53}$ to about $10^{19}$~GeV, modify the running
of the quartic coupling, and avoid the appearance of problems in known physics. In the following we
show a model where we achieve the objectives listed above.  
The aim of this exercise is to show a reasonable possibility where a ULP with the required 
properties manifests itself. While we do not claim that our solution is unique, we want to stress at the 
beginning that finding a scenario which is not in contradiction with known physics, and at the same time fulfills 
some requirements (specified below) is not easy. Furthermore, if we require that the new scenario solves 
the instability of the Higgs potential then the possibilities are reduced to basically one scenario.

The possibility that a Landau pole may be present 
at the Planck mass is not totally new. 
The authors of~\cite{Maiani:1977cg} used this hypothesis to set bounds for the 
number and masses of quarks and leptons. Later the possibility of an unification at strong coupling has been 
studied in the context of GUT and SUSY, see e.g.~\cite{Rubakov, Liu}, and references therein for a review.  We follow a
different line:  ULP is related with physics at the
Planck scale and we do not introduce a new gauge group nor SUSY, and no new gauge or Higgs fields. 
Another interesting proposal, which has some similarity with ours,
relies on the possible existence of a non-gaussian UV fixed point. This conjecture arises from the possibility
of gravity being an asymptotically safe theory~\cite{asymsafe} and assuming that all other interactions unify 
at the same nonperturbative fixed point (see recent discussion in \cite{shaposh}) and, remarkably
enough, this conjecture may have interesting implications on the quartic coupling as well. 
This latter proposal does not assume a radical 
new theory for gravity --just an hypothetical nonperturbative completion of the usual theory.

{}From now on we will adhere to the minimal hypothesis of the existence of a ULP.
In this letter we will present our calculations at the one loop level. Present theoretical knowledge would 
in principle enable us to perform the calculation up to three loops~\cite{3loops} and we will present 
a full analysis of the two-loops calculation in a forthcoming work~\cite{inpreparation}.  The two-loop
result will not significantly be different from the one presented here because we perform the analysis 
in the region below the first new threshold (where the top Yukawa coupling is largest $\sim 1$ and perturbation
theory in the scalar sector worst) with the full two loops theory, and used the one-loop approximation only
above this first threshold where $Y_t$ is smaller.

At one loop the running of the couplings is given by simple equations
\be
\frac{d g_{i}(t)}{d t} = \beta_{i}(t),\quad \beta_{i}\equiv \frac{1}{16\pi^2}g_{i}^3 b_{i}, 
\quad t\equiv\log{\frac{\mu}{GeV}},\la{RG}
\ee
where $i=1,2,3$ represents the U(1), SU(2) and SU(3) couplings respectively.
The presence of new particles will alter this running. At one loop the behavior of 
\be
\frac{1}{\alpha_i}\equiv\frac{4\pi}{g_i^2}
\ee
are linear. The presence of new particles just alters the slope of straight lines.

We now present our model, explicitly spelling out the constraints we impose.
We require the model to have the Standard Model gauge group and particles coming in generations with 
the same quantum numbers as the usual ones. This ensures, for example,  that there will be color singlets made 
of three fermions.  We will assume that the various particles have masses such that they contribute 
only when a particular threshold of energy is reached. The full evolution is therefore given by a set 
of straight segments and the solution is found matching the boundary conditions. 
 To differentiate the new from the old particles
we will call them ``quarkons'' and ``leptos''. To avoid problems with anomalies, and the introduction 
of new Higgs-like particles, all new particles are {\it vector-like} Dirac particles, 
 but they maintain the representations 
of the known gauge groups. This also avoids that the presence of new particles coupling to the Higgs could 
actually render the instability of the quartic term more pronounced. In fact recent LHC results \cite{nofourth} 
severely constrain the possibility of having a (Standard Model like) fourth generation coupled to the Higgs particle, 
no matter how large its mass. However, new heavy fermions appearing 
in vector representations are still largely unconstrained.   

We emphasize that all quarkons and 
leptos are {\it vector-like} particles and that they do not get their masses through the Higgs mechanism. 
In particular one can consider them to have Dirac masses. Thus the Higgs boson Lagrangian has 
the form of SM and therefore the one-loop RG flow of quartic coupling and of Yukawa couplings 
are fully controlled by the behavior of gauge couplings.
Baryon and lepton charges for new fermions are conserved separately.
Quarkons are SU(3) triplets and leptos are SU(3) singlets. They both come in two kinds: SU(2) doublets,
which we will call L-quarkons and L-leptos,  with the hypercharges of left quarks and left leptons respectively;
and SU(2) singlets, R-quarkons, with the hypercharge of right handed quarks, and R-leptos 
carrying the hypercharge of right electrons. 
However, R-leptos which are singlets for all SM groups (like right neutrinos) could be 
 present in principle, but since they do not contribute to the running of the gauge
couplings we can ignore them. 

For the $U(1)$ gauge coupling $g_1$, the constant $b_1$ is given by:
\bea
b_1 &=& \frac{41}{6} + \frac{2}{3}N_{\mathbf{L-leptos}} + \frac{4}{3}N_{\mathbf{R-leptos}} \nonumber \\
&& + \frac{2}{9}N_{\mathbf{L-quarkon}} +\frac{20}{9}N_{\mathbf{R-quarkon}}
. \la{b1}
\eea
For the $SU(2)$ gauge coupling we have:
\be
b_2  = -\frac{19}{6} + \frac{2}{3}N_{\mathbf{L-leptos}} + 2 N_{\mathbf{L-quarkon}}
.\la{b2}
\ee
For the $SU(3)$ gauge coupling
\be
b_3 = -7 +\frac{4}{3}\left(N_{\mathbf{L-quarkon}}+N_{\mathbf{R-quarkon}}\right). \la{b3}
\ee
The integers $N$ in these formulas refer to the number of quarkon and leptos
multiplets contributing to beta functions.

Since the coefficients are piecewise constant, and change at the energies representing 
the scale at which the new particles appear, it is possible to do a systematic search. We have imposed as a
boundary condition of the differential equation that $1/\alpha_i=0$ at the Planck scale $m_p$. 
In any case the model cannot be trusted at energies approaching $m_p$ for more than one reason. 
The perturbative approach will have broken down, not to speak of the one loop approximation, 
and moreover gravitational effects could not be ignored. Our setting a precise boundary condition 
giving a common pole at a particular scale is therefore just expedient to describe a common pole 
that the present theoretical tools cannot properly describe.

The other low energy boundary conditions are given by the experimental values:
$\alpha_s=0.1184$, $g_1=0.358729$, $g_2=0.648382$, $g_3=1.16471$, $y=0.937982$, $\lambda=0.125769$
for $M_H=124 GeV$ at the scale of the top mass $\mu=M_t=172.9 GeV$. These values are 
insensitive to $M_H$ in the range $124-126$~GeV. Since the equations are linear (at one loop), 
the results presented here are quite ``robust'' for slight changes in the boundary condition, 
both in the low energy and pole regions.

In principle there would be four kinds of particles that switch on at four scales, and the boundary 
conditions at the intermediate scales impose three constraints. We require the scales be 
between the TeV region and the ULP, and that the evolution is monotonic (the curves must not intersect themselves). 
This results in the only allowed order of the different thresholds as one goes up being the following: 
L-quarkons, R-quarkons, L-Leptos, R-Leptos. If one requires the switching on of the leptos to be at 
the same scale one finds solutions. On the contrary setting the quarkons at the same scale does not provide a solution. 
This enables us to reduce the number of parameters to three, with three equations, and therefore find 
a unique solution. Since the scale for the leptos must be larger than the one of quarkons and therefore closer
to the Planck scale the possibility of splitting the two scales of the leptos gives just a little uncertainty 
at very high energies. We were also able to fix the number of generations. For three generations or 
less there is no solution in the physical range. For more than four generations the quartic coupling 
develops an instability. 

Therefore, we consider 
that there are four identical ``generations''. 
With the given constraints the particles must be at  the following scales:
\begin{itemize}
\item At  $5.0~10^3$ GeV the L-quarkons ($N_{\mathbf{L-quarkon}}=4$).

\item At $3.7~10^{7}$ GeV the R-quarkons ($N_{\mathbf{R-quarkon}}=4$).

\item At $2.6~10^{14}$ GeV the L and R-leptos ($N_{\mathbf{L-leptos}}=N_{\mathbf{R-leptos}}=4$).

\end{itemize}
It is interesting to remark that the lowest threshold in this minimal ULP model is
tantalizingly close to the reach of the LHC but certainly not excluded by present data.

\begin{figure}[htb]
\includegraphics[scale=.45]
{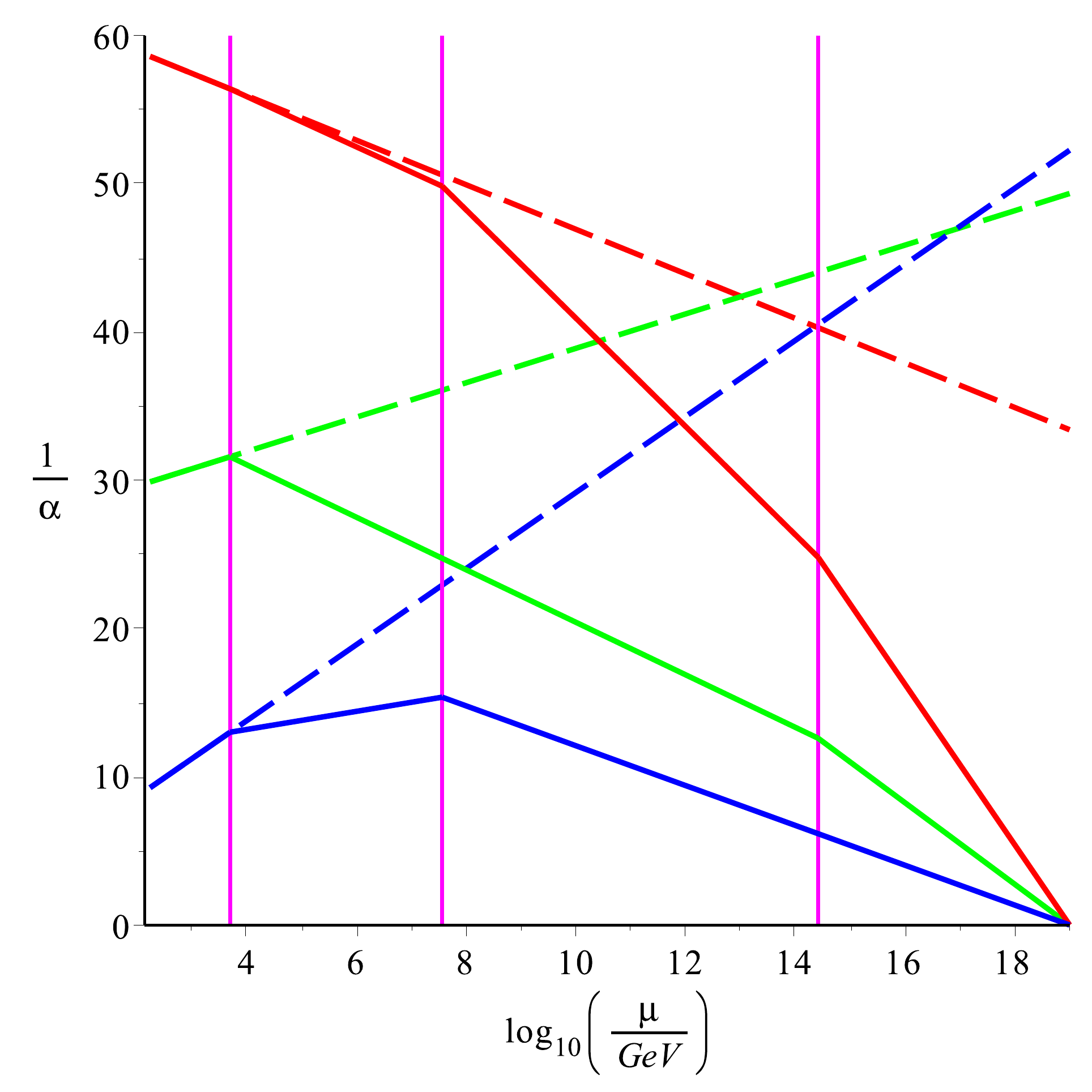}

\caption{\sl The running of $\alpha_i$, the inverse of the gauge couplings. The dotted lines are the runnings
 in the absence 
of quarkons and leptos. The $\alpha_i$ are in descending order as $i$ increases.} \label{gaugerunfig}
\end{figure}

In Fig.~\ref{gaugerunfig}
we show the running of the gauge coupling. (the initial running shown is actually made with the two-loop equation).  
One can see that the hierarchy of the couplings is respected, the strong force remains stronger than 
the weak. The scale at which there is the appearance of the new particles is larger than the experimental 
bounds on the presence of new fermions, but not too much. This scenario shows that the ULP may exist with 
new physics at energies within reach.  Other solutions are possible and we 
will comment on them in~\cite{inpreparation}, although the qualitative features of these 
alternative options are similar to the one presented here.

The running of the gauge couplings affects the other couplings we considered. 
As far the top Yukawa coupling is concerned the equation is
\be
\beta_y^{(1)} = \frac{1}{\left(4\pi\right)^2}\,{ y}\, \left( -\frac{9}{4}\,{{ g_2}}^{2}-{\frac {17}{12}}\,{{
g_1}}^{2}-8\,{{ g_3}}^{2}+\frac{9}{2}\,{{ y}}^{2} \right).
\ee
\begin{figure}[htb]
\includegraphics[scale=.45]
{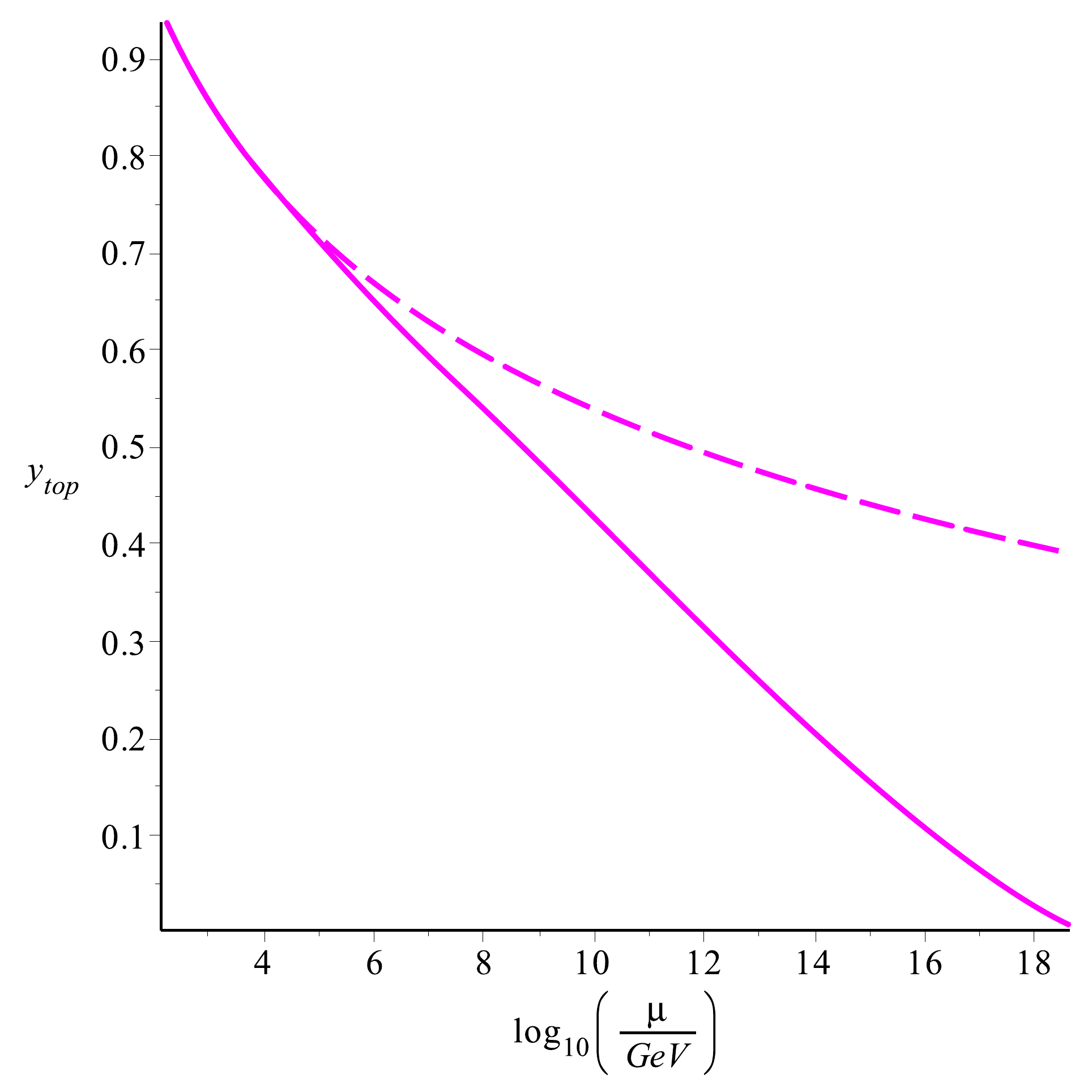}
\caption{\sl The running of $y_t$. The dotted line is the SM case.}  \label{yrunfig}
\end{figure}
Solving the differential equation we obtain Fig.~\ref{yrunfig}. The coupling is undistinguishable 
from the standard model for energies up to $10^6$~GeV, and vanishes at the ULP.

\begin{figure}[htb]
\includegraphics[scale=.45]
{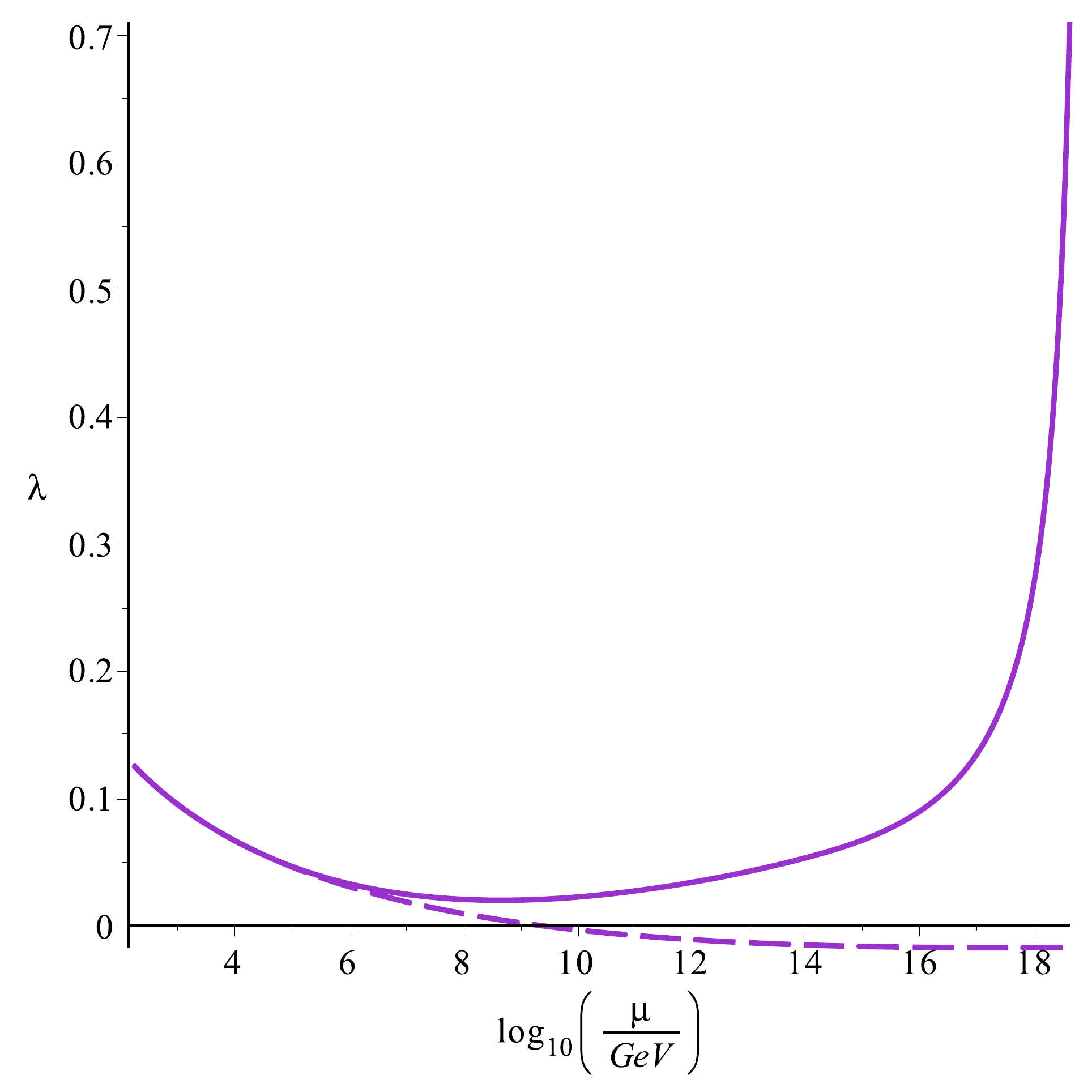}
\caption{\sl The running of quartic coupling of the Higgs field. 
The dotted line shows the instability that the standard model develops in the presence of a 
Higgs mass of 124-126 GeV.}  \label{lambdarunfig}
\end{figure}

The quartic coupling equation is
\begin{eqnarray}
\beta_{\lambda}^{(1)} &=& \frac{1}{16\pi^2}\left( 24\,{\lambda}^{2}-6\,{y}^{4}+\frac{3}{4}\,g_2^4+\frac{3}{8}\,
 \left( {g_{{2}}}^{2}+{g_{{1}}}^{2} \right) ^{2}
\right.\nonumber\\  && \left.
+ \left( -9\,{g_{{2}}}
^{2}-3\,{g_{{1}}}^{2}+12\,{y}^{2} \right) \lambda\right).
\end{eqnarray}
Its solution, given the couplings we calculated earlier, is shown in Fig.~\ref{lambdarunfig}.
We see that the quartic coupling for our choice of new particles comes close to vanishing, but 
never actually becomes negative. At the ULP the coupling diverges. Therefore, the potential term 
will dominate over the other terms. The mass parameter remains finite with a regular behavior 
while the evolution is in the perturbative regime.

We expect that taking into account two-loop corrections does not
affect  the position of the successive thresholds dramatically. Indeed, from the SM we know,
that the gauge running remains almost unchanged after taking into account
of two and even three loops: the renormalization group trajectories almost coincide with the lines of
constant slope obtained in a one loop computation \cite{3loops}.
Since all our scales lie in a perturbative regime one expects to have the
same property in the ULP scenario. Thus we conclude that one can
trust the one loop determination of the  masses of quarkons and leptos.

The scenario described in this letter may give hints to aspects of the trans-planckian phase where the 
kinetic term becomes negligible and the propagator ``freezes'', and the Higgs 
(which might be composite) decouples. This suggests that gauge bosons may possibly be 
effective (probably composite too) degrees of freedom. 
Gravity will play a dominant role, but the 
absence of propagating degrees of freedom suggests a ``geometry'' without points, with space-time 
possibly described by a noncommutative geometry~\cite{ncg} and/or replaced by a pregeometric entity, such 
as the spin foam and spin networks of quantum gravity~\cite{rovelli}. In proposals 
such as \cite{AlfaroEspriuPuigdomenech}, even gravity
is described entirely in terms of fermions, which may describe the whole physical 
world if a scenario like an ULP is realized.

It could well be the case that the onset of gravity corrections renders the ULP we advocate
in this paper non-singular. Indeed gravity being non-renormalizable will require higher-dimensional 
operators with more derivatives to render the theory finite. In particular, we expect dimension six kinetic terms like
 \be
 \frac{\gamma}{2M^2_P} \mbox{tr}\left(D_\mu W^{\mu\nu}D_\mu W^{\mu}_{\nu}\right) + \cdots.
 \ee
This would correspond to a renormalization of the gauge coupling induced by gravity of the form
 \be
 \frac{1}{g^2(p^2)} \simeq \beta_0 \log\frac{m^2_P}{p^2} + \gamma \frac{p^2}{m^2_P}.
 \ee
As shown in \cite{shaposh} gravitational corrections may drive the ULP towards a new fixed point. 

The renormalization flow of the various constants, especially in view of the new data coming from the LHC, 
may be an essential tool for the understanding of physics at the gravitational frontier.

\begin{acknowledgments}
We thank  M.~Zoller for a private communication regarding the choice of initial conditions.
A.A., D.E. and F.L. are partially supported by projects FPA2010-20807, 2009SGR502 and CPAN (Consolider CSD2007-00042). 
This work has been made possible by the INFN-MCINN bilateral agreement AIC-D-2011-0815.
A.A.A. and M.A.K. are also supported by Grant RFBR 13-02-00127 and by SPbSU grant 11.38.660.2013.
\end{acknowledgments}


\begin{thebibliography}{99}

\bibitem{particledata} J. Beringer et al. (Particle Data Group), Phys. Rev. D86, 010001 (2012).

\bibitem{landaupole}  L. D. Landau, A. A. Abrikosov, and I. M. Khalatnikov, Dokl. Akad.
Nauk SSSR 95, 497, 773, 1177 (1954);
 L.D.Landau, I.Ya.Pomeranchuk, Dokl. Akad. Nauk SSSR 102, 489 (1955).


\bibitem{HiggsMass} G. Aad et al. [The ATLAS collaboration], Phys. Lett. B 716 (2012) 1; S.Chatrchyan 
et al. [The CMS collaboration], Phys. Lett. B 716 (2012) 30.

\bibitem{AlfaroEspriuPuigdomenech}
J.~Alfaro, D.~Espriu and D.~Puigdomenech,
  ``Spontaneous generation of geometry in four dimensions,''
  Phys.\ Rev.\ D {\bf 86}, 025015 (2012)
  [arXiv:1201.4697 [hep-th]].

\bibitem{instability} N. V. Krasnikov,  Yad. Fiz. 28 (1978) 549;
P.Q. Hung, Phys. Rev. Lett. 42 (1979) 873;
H.D. Politzer and S. Wolfram,
Phys. Lett. B 82 (1979) 242 [Erratum ibid. 83B (1979) 421].

\bibitem{EliasMiro:2011aa}
  J.~Elias-Miro, J.~R.~Espinosa, G.~F.~Giudice, G.~Isidori, A.~Riotto and A.~Strumia,
  ``Higgs mass implications on the stability of the electroweak vacuum,''
  Phys.\ Lett.\ B {\bf 709}, 222 (2012)
  [arXiv:1112.3022 [hep-ph]];
  F.~Bezrukov, M.~Y.~.Kalmykov, B.~A.~Kniehl and M.~Shaposhnikov,
  ``Higgs Boson Mass and New Physics,''
  JHEP {\bf 1210}, 140 (2012).




\bibitem{Maiani:1977cg}
  L.~Maiani, G.~Parisi and R.~Petronzio,
  ``Bounds on the Number and Masses of Quarks and Leptons,''
  Nucl.\ Phys.\ B {\bf 136}, 115 (1978).

  
\bibitem{Rubakov}
V.~A.~Rubakov and S.~V.~Troitsky,
  ``Trends in grand unification: Unification at strong coupling and composite models,''
  hep-ph/0001213.


\bibitem{Liu} C.~Liu,
  ``[SU(3) x SU(2) x U(1)]**2 and strong unification,''
  Phys.\ Lett.\ B {\bf 591}, 137 (2004)
  [hep-ph/0405271].


\bibitem{asymsafe} S. Weinberg, Ultraviolet divergences in quantum theories of gravitation,â in: General 
 Relativity. (S. W. Hawking and W. Israel, eds.), 
 Cambridge Univ. Press, Cambridge (1979), p. 790 - 831;\ M. Niedermaier and M. Reuter, 
 Living Rev. Rel., 9, 2006-5 (2006).

\bibitem{shaposh}
M.~Shaposhnikov and C.~Wetterich,
  ``Asymptotic safety of gravity and the Higgs boson mass,''
  Phys.\ Lett.\ B {\bf 683}, 196 (2010)
  [arXiv:0912.0208 [hep-th]];
  M.~E.~Shaposhnikov,
  ``Asymptotic safety of gravity and the Higgs-boson mass,''
  Theor.\ Math.\ Phys.\  {\bf 170}, 229 (2012)
  [Teor.\ Mat.\ Fiz.\  {\bf 170}, 280 (2012)].



\bibitem{3loops}K.~G.~Chetyrkin and M.~F.~Zoller,
  ``Three-loop $\beta$-functions for top-Yukawa and the Higgs self-interaction in the Standard Model,''
  JHEP {\bf 1206}, 033 (2012)
  [arXiv:1205.2892 [hep-ph]].

\bibitem{inpreparation} A.A.~Andrianov. D.~Espriu, M.A.~Kurkov and F.~Lizzi, in preparation.

\bibitem{nofourth} See e.g. A. Lenz, Advances in High Energy Physics 2013 (2013), 910275,
http://dx.doi.org/10.1155/2013/910275


\bibitem{ncg}  A. Connes, M. Marcolli, ``Noncommutative Geometry, Quantum Fields and Motives'', AMS 2007; P.~Aschieri, M.~Dimitrijevich, P. Kulish, F.~Lizzi, J.~Wess, ``Noncommutative Spacetimes'', Springer Lecture Notes 774, 2009.

\bibitem{rovelli} C. Rovelli, ``Quantum Gravity'', Cambridge University Press, 2007.




\end{thebibliography}
\end{document}